\documentclass[12pt]{article}
\usepackage{setspace}
\usepackage{epsfig}
\usepackage{vmargin}
\begin{document}
\setpapersize{USletter}
\setmarginsrb{1.25in}{1in}{1in}{1in}{0pt}{0mm}{0pt}{36pt}
\begin{spacing}{1.5}
\renewcommand{\baselinestretch}{1.5}

\vspace{1.5in}
\begin{center}

{\sffamily\bfseries\LARGE
\noindent
Are the Bader Laplacian and the \\
Bohm Quantum Potential Equivalent?
\\
}
\vspace{0.5in}
{\sffamily
{\bfseries
\noindent	
  {Creon Levit$^{*}$ \& Jack Sarfatti$^{\dag}$} \\ 
}
\noindent	
  {\small $^{*}$NASA Ames Research Center, creon@nas.nasa.gov}\\
\noindent	
  {\small $^{\dag}$Internet Science Eductaion Project,  sarfatti@well.com}\\
}
\end{center}

\vspace {.25in}

{\bfseries
\begin{center} 
ABSTRACT \\
\end{center}
The de Broglie-Bohm ontological interpretation of quantum
theory\cite{bohm,holland} clarifies the understanding of many
otherwise counter-intuitive quantum mechanical phenomena.  We report
here on an application of Bohm's quantum potential to the bonding and
reactivity of small molecules.  In the field of quantum chemistry,
Bader has shown\cite{bader} that the topology of the Laplacian of the
electronic charge density characterizes many features of molecular
structure and reactivity.  Examination of high accuracy {\em
ab-initio} solutions for several small molecules suggests that the
Laplacian of Bader and the quantum potential of Bohm are structurally
equivalent.  It is possible that Bohmian mechanics using the quantum
potential can make quantum chemistry as clear as it makes
non-relativistic quantum mechanics.

}

\pagebreak

A set of {\em ab-initio} computational experiments were performed
simulating H$_{2}$O, H$_{2}$O$_{2}$, and C$_{2}$H$_{4}$ to obtain the
Laplacian of the electronic charge density ($\nabla^2\rho$).  Bader has
shown\cite{bader} that $\nabla^{2}\rho$ localizes lone pairs, bonded
charge concentrations and regions subject to electrophilic or
nucleophilic attack.  $\nabla^{2}\rho$ acts as an objective {\em
electron localization function}\cite{becke,silvi,bader2,kulkarni}, and is
free from the difficulties encountered when analyzing a wavefunction
by decomposition into a particular orbital basis.

In order to explore the application of the de Broglie--Bohm
interpretation of quantum mechanics\cite{bohm,holland} to molecular
physics, we also calculated the Bohm quantum potential $Q$ for the
same molecules.  Visual comparison of the three dimensional structure
of these two fields ($Q$ and $\nabla^{2}\rho$) strongly suggests they
are equivalent.

Density functional theory calculations were carried out for the
optimized ground state equilibrium geometries of water, hydrogen
peroxide, and ethylene using Gaussian 94\cite{g94} at the {\sc
b3lyp}/6-311++{\sc g}(2d,2p) level of theory\cite {b3lyp}.  Isosurfaces of
the Laplacian of the electronic charge density, $\nabla^{2}\rho=C$,
were visualized\cite{fast} for these molecules over a range of values
of $C$ and are shown in figures 1-4 for the case of
water\cite{supplementary}.

Following the analysis of Bader\cite{bader}, regions of
$\nabla^{2}\rho<0$ correspond to electronic charge concentration, and
conversely regions of $\nabla^{2}\rho>0$ correspond to regions of
charge depletion.  The four views of H$_{2}$O shown in Figure 1
correspond to a very negative value of $\nabla^{2}\rho$. The
electronic charge concentrations associated with the nuclei as well as
the two lone pair charge concentrations on the oxygen atom are
visible.  For a less negative value of the Laplacian (figure 2),
concentrations of electronic charge on the OH bonds become visible.
At $\nabla^{2}\rho=0$ (figure 3) the outline of the entire molecule
emerges.  Finally, for a positive value of $\nabla^{2}\rho$ (figure
4), three regions of valence charge depletion are visible, as well as
the region of depletion surrounding the Oxygen's core electrons.

The structure of $\nabla^{2}\rho$ may be formally defined in terms of
the attractors, repellors, and saddle points in $\nabla\nabla^{2}\rho$
and their interconnections\cite{bader}.  In the case of water, there
is one attractor at each nucleus, one in each lone pair region, one on
each bond, and no others (figures 1-3).

To compute Bohm's quantum potential\cite{bohm} we first express the
wavefunction in polar form, $\Psi=Re^{iS/\hbar}$.  The quantum
potential for an $N$ electron molecule using the Born-Oppenheimer
approximation is then:
\[Q=-\frac{\hbar^{2}}{2m_{e}}\sum_{i=1}^{N}\frac{\nabla_{i}^{2}R}{R}.\]
$Q$ is a function of the $3N$ coordinates of all of the electrons, and
$\nabla Q$ is, in Bohm's interpretation of quantum mechanics, the
nonlocal quantum force that acts in addition to the classical forces
on each electron.

For a one-electron molecule, $Q$ reduces to
\[Q_{1}=-\frac{\hbar^{2}}{2m_{e}}\frac{\nabla^2\sqrt{\rho}}{\sqrt{\rho}}.\] 
Visualization in three-dimensional physical space of the $3N$
dimensional quantum potential for multiple electron molecules was
accomplished by calculating ${\overline Q}$ as a function of
$\Psi_{1}=\sqrt{\rho/N}$ where $\Psi_{1}$ is the solution of the exact
one-electron Schr\"odinger equation for an $N$-electron
molecule\cite{hunter}.  In this case, ${\overline Q}$ is identical to
$Q_{1}$ times a constant.  In terms of the charge density $(\rho)$ and
the Bader Laplacian $(\nabla^{2}\rho)$,
\[{\overline Q}=\nabla^{2}\rho/2\rho-(\nabla\rho/2\rho)^{2}.\]

Isosurfaces of the (one electron) Bohm quantum potential for H$_{2}$O
are shown in figures 5-8 for a range of values.  The topology of this
field (the structure of its foliation by isosurfaces, or,
equivalently, the attractor, repellor, and saddle connections of its
gradient) appears identical to that of $\nabla^{2}\rho$ in the case of
water, as well as in the cases of all other molecules examined by the
authors.

There are some geometrical differences between the two fields.  The
presence of the factor $\rho^{-1/2}$ in the expression for $\overline
Q$ implies that the magnitude of $\overline Q$ may be large far from
the nuclei.  In Bohm's picture, this factor accounts for nonlocality
(distance independence) in quantum mechanics.  In the molecules we
examined, this factor accounts for structures in $\overline Q$
extending over larger distances than those in $\nabla^{2}\rho$.  It
may also be that the quantum force $\nabla Q$ acting on the electrons
is responsible, in part, for the ``concentration'' of electronic
charge (i.e. regions of Bader Laplacian $<0$) in bonding and lone pair
locations.

While other authors have compared the structure of $\nabla^{2}\rho$
with $\nabla^{2}\sqrt{\rho}/\sqrt{\rho}$ for atoms\cite{kohout} and
molecules\cite{hunter2}, this is the first time it has been done
in three dimensions, and to our knowledge the first time that a
connection of the Bader chemical theory to the de Broglie--Bohm quantum
potential has been made.

\pagebreak

\end{spacing}


\pagebreak
\setmarginsrb{0.25in}{1in}{0.25in}{1in}{0pt}{0mm}{0pt}{0mm}
\noindent

\begin{tabular}{ p{3.65in} p{3.65in}}
\epsfig{figure=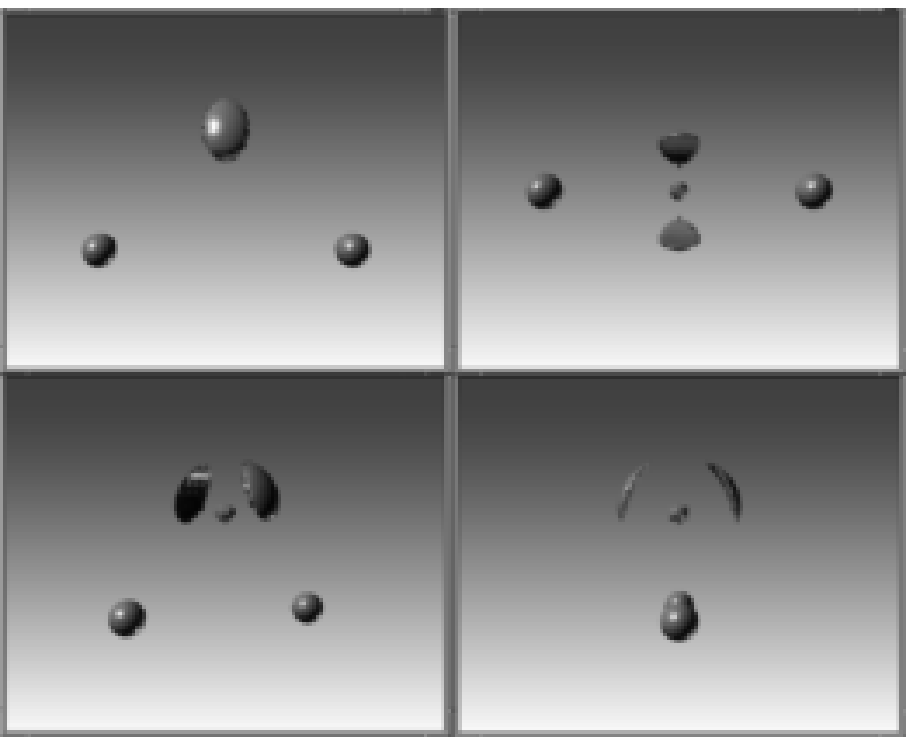} &
\epsfig{figure=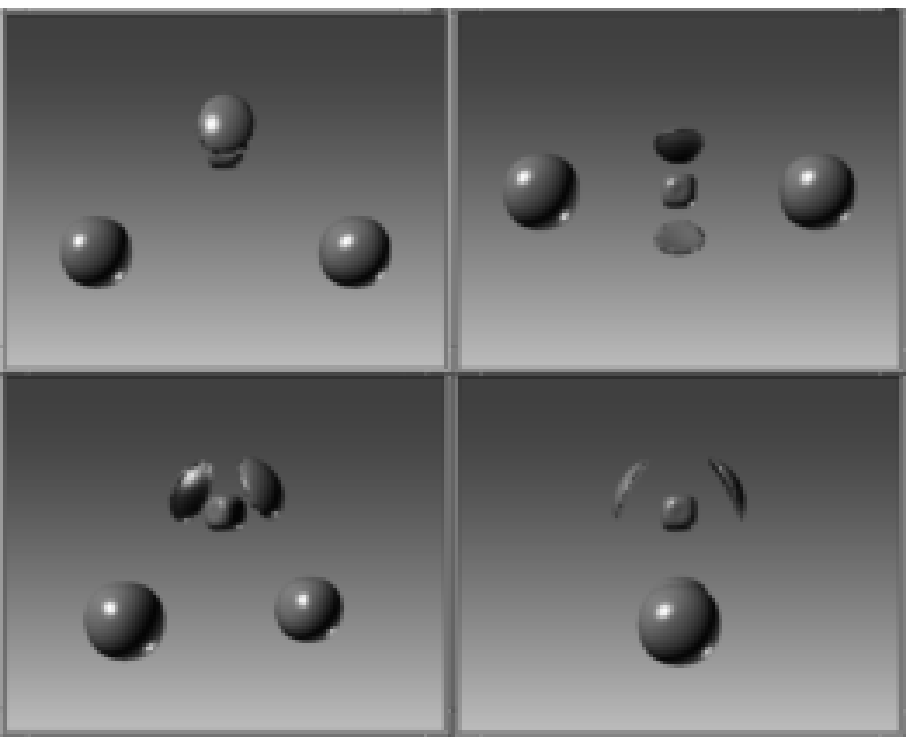} \\
figure 1 & figure 5 \\
\vspace {1in} \\
\epsfig{figure=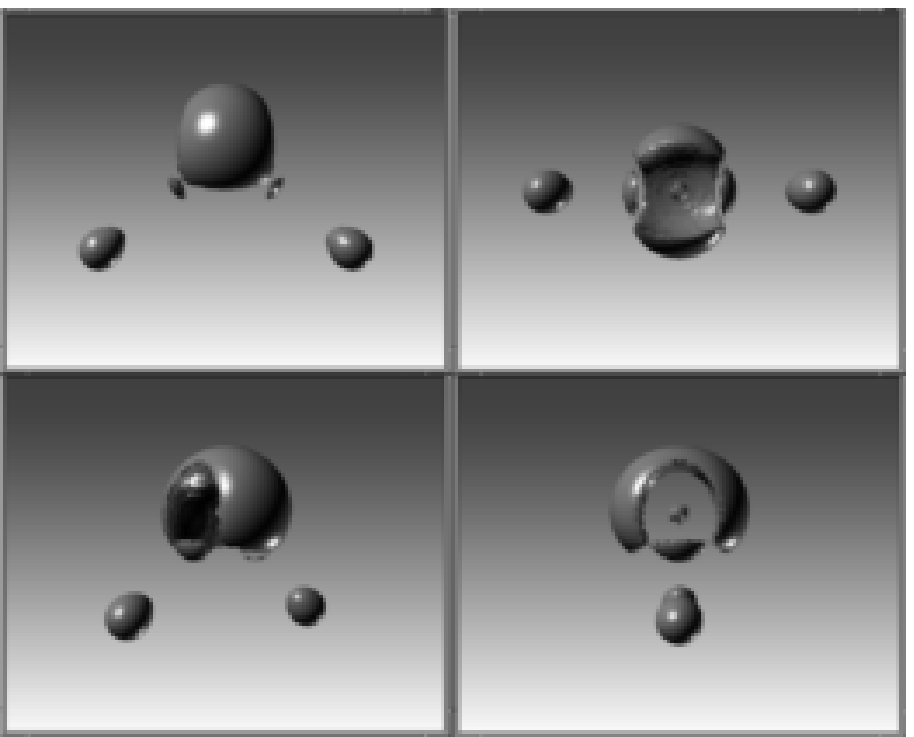} &
\epsfig{figure=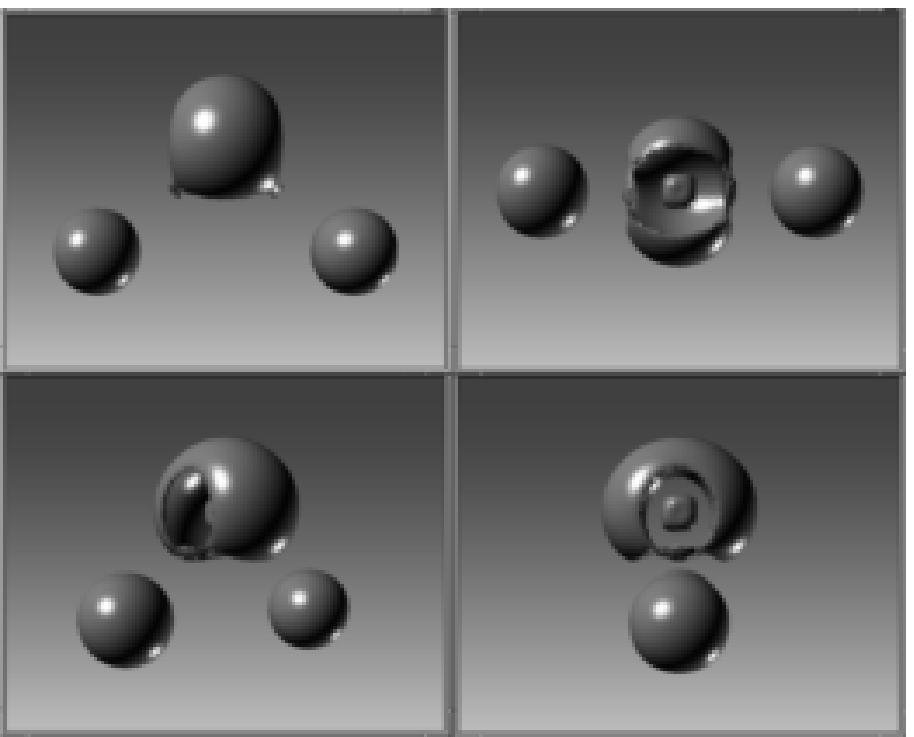} \\
figure 2 & figure 6 \\
\end{tabular}

\pagebreak

\begin{tabular}{ p{3.65in} p{3.65in}}
\epsfig{figure=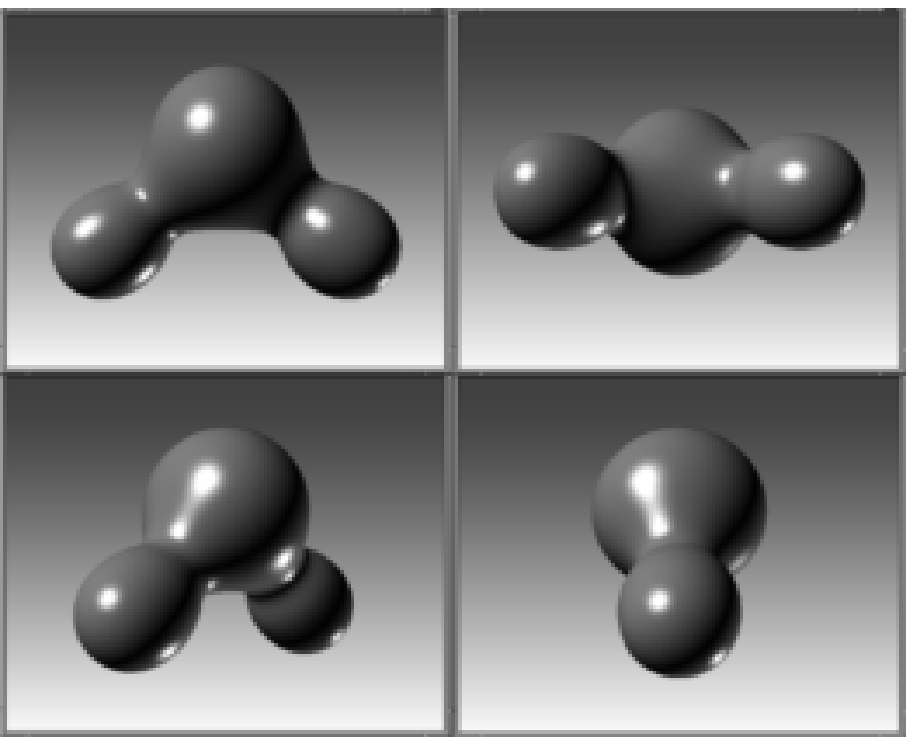} &
\epsfig{figure=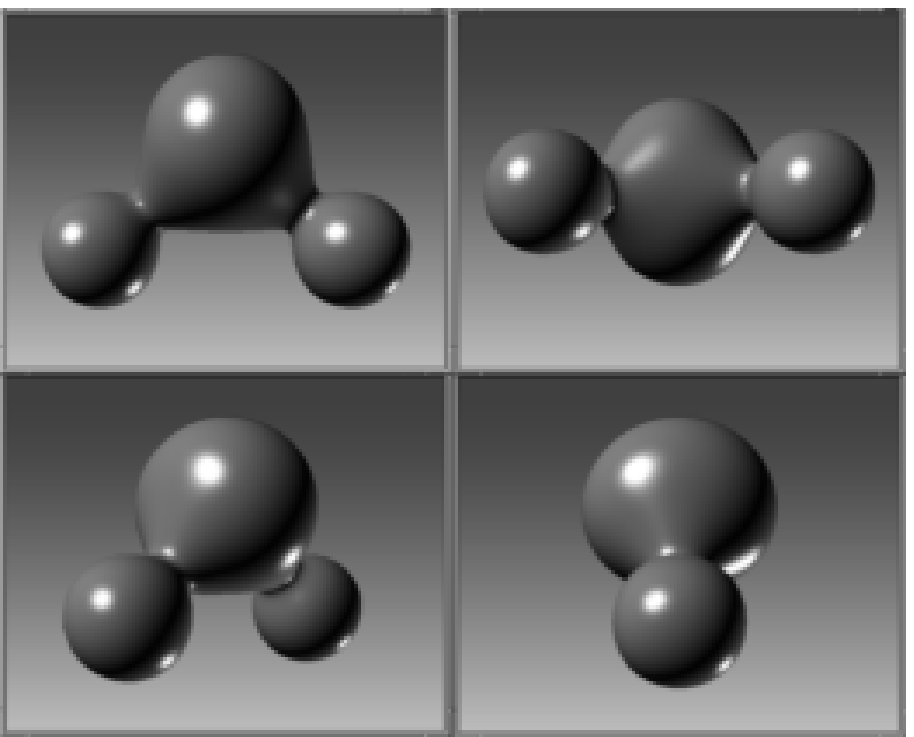} \\
figure 3 & figure 7 \\
\vspace {.35in} \\
\epsfig{figure=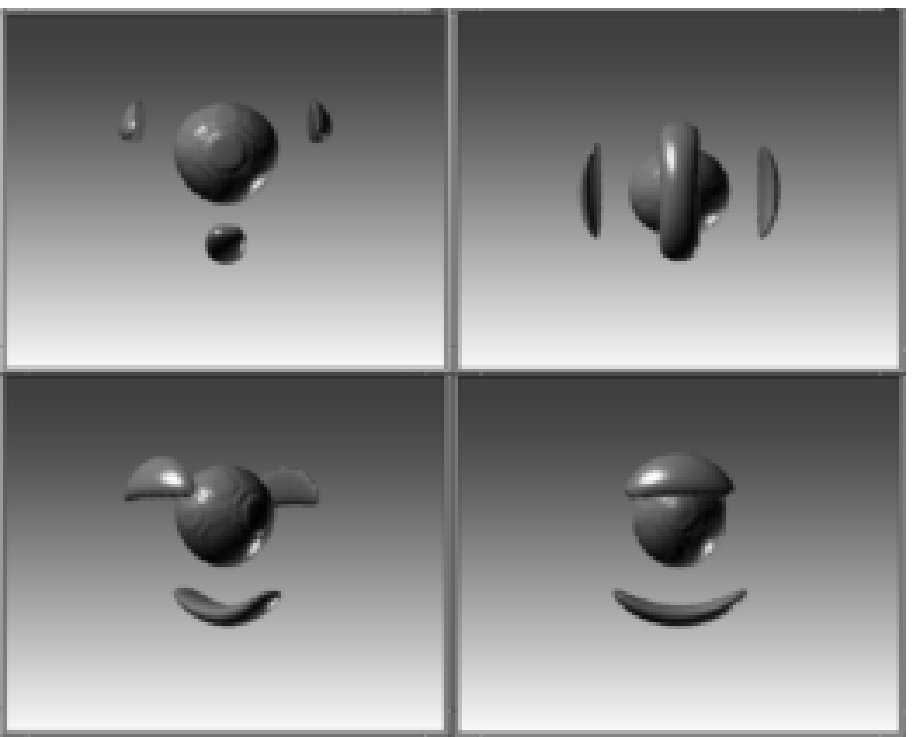} &
\epsfig{figure=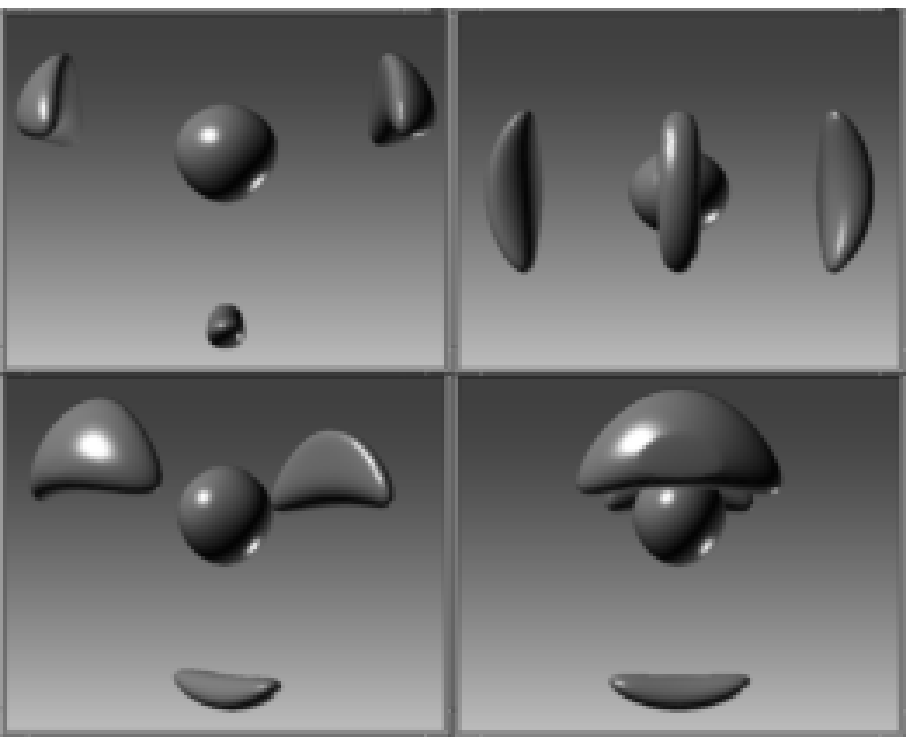} \\
figure 4 & figure 8 \\

\vspace {0.2in} \\
{\small
figures 1-4: Each figure shows four views of an isoscalar surface of
the Laplacian of the charge density for H$_{2}$O.  Figure 1
($\nabla^2\rho \ll 0$) shows local minima at the nuclei and at the
locations of the oxygen lone pairs.  Figure 2 ($\nabla^2\rho < 0$)
shows additional local minima on the bonds.  In figure 3
($\nabla^2\rho = 0$) the outline of the entire molecule is visible.
Figure 4 ($\nabla^2\rho > 0$) shows regions of valence charge
depletion. } &
{\small
Figures 5-8: Each figure shows four views of an isoscalar surface of
the (one electron) Bohm quantum potential for H$_2$O. The topology and
symmetry are identical to the Bader Laplacian ($\nabla^2\rho$,
figs. 1-4) for H$_2$O and other molecules.
}
\\

\end{tabular}

\pagebreak
\setmarginsrb{1.25in}{1in}{1in}{1in}{0pt}{0mm}{0pt}{36pt}
{\sffamily\footnotesize
\noindent
{Correspondence and requests for materials to: 
\\
Creon Levit\\
MS T27A\\
NASA Ames Research Center\\
Moffett Field, CA.  \\
USA\\
creon@nas.nasa.gov\\
}
}
\end{document}